\documentclass{aa} 
\usepackage{graphicx}
\usepackage{txfonts}
\usepackage{natbib}
\usepackage{url}
\bibpunct{(}{)}{;}{a}{}{,}
\usepackage[breaklinks=true]{hyperref} %% to avoid \citeads line fills
\hypersetup{colorlinks=true,urlcolor=blue,citecolor=blue,pdfborder= 0 0 0}

\newcommand{\Mpc}{$h^{-1}$\thinspace Mpc}

\newcommand{\vmh}{h^{-1}\mathrm{Mpc} }

\begin{document}

\title{BOSS Great Wall: morphology, luminosity, and mass}

\author{
Maret Einasto \inst{1} \and
Heidi Lietzen \inst{2} \and
Mirt Gramann \inst{1} \and
Enn Saar \inst{1} \and
Elmo Tempel \inst{1,3} \and
Lauri Juhan Liivam\"agi \inst{1,4} \and
Antonio D. Montero-Dorta \inst{5} \and
Alina Streblyanska \inst{6,7}\and
Claudia Maraston \inst{8}\and
José Alberto Rubiño-Martín \inst{6,7}
}

\institute{  
Tartu Observatory, Observatooriumi 1, 61602 T\~oravere, Estonia
  \and
Tuorla Observatory, Department of Physics and Astronomy, University of Turku, 
V\"ais\"al\"antie 20, 21500 Piikki\"o, Finland
\and
Leibniz-Institut f\"ur Astrophysik Potsdam (AIP), An der Sternwarte 16, D-14482
Potsdam, Germany
\and
Institute of Physics, University of Tartu, W.Ostwaldi 1, 50411 Tartu, Estonia
\and
Department of Physics and Astronomy, The University of Utah, 115 South 1400 East, Salt Lake City, UT 84112, USA
\and
Instituto de Astrofísica de Canarias, E-38200 La Laguna, Tenerife, Spain
\and
Universidad de La Laguna, Dept. Astrofísica, E-38206 La Laguna, Tenerife, Spain
\and
ICG-University of Portsmouth, Dennis Sciama Building, Burnaby Road, PO1 3FX, Portsmouth, United Kingdom
    }

\abstract
{
Galaxy superclusters are the largest systems in the Universe 
that can give us information about the formation and evolution of 
the cosmic web. 
}
{
We study the morphology of the superclusters from the 
BOSS Great Wall (BGW), a recently discovered very rich supercluster complex 
at the redshift  $z = 0.47$.
}
{
We have employed the Minkowski functionals to quantify supercluster morphology.
We calculate supercluster luminosities and masses using two methods.
Firstly, we used data
about the luminosities and stellar masses of high stellar mass galaxies with
$\log(M_*/h^{-1}M_\odot) \geq 11.3$.
Secondly, we applied a scaling relation that combines morphological
and physical parameters of superclusters 
to obtain supercluster luminosities, and 
obtained supercluster masses using the mass-to-light ratios found
for local rich superclusters.
}
{
The BGW superclusters are very elongated  systems,
with shape parameter values of less than $0.2$.
This value is lower  than that found for the most elongated
local superclusters. 
The values of the fourth Minkowski functional $V_3$ for the 
richer BGW superclusters ($V_3 = 7$ and $10$) show that they have 
a complicated and rich inner structure. 
We identify  
two Planck SZ clusters in the BGW superclusters, one in the richest BGW supercluster, 
and another in one of the poor BGW superclusters.
The luminosities of the
BGW superclusters are in the range of $1 - 8\times10^{13}h^{-2} L_{\sun}$,
and masses in the range of
 $0.4 - 2.1\times~10^{16}h^{-1}M_\odot$. 
Supercluster luminosities and masses obtained with two methods agree well.
}
{
The BGW is a complex of massive, luminous and large superclusters
with very elongated shape. The search and detailed study, including the
morphology analysis of the richest superclusters and their complexes from
observations and simulations can help us to understand formation
and evolution of the cosmic web.
}

\keywords{Large-scale structure of the Universe - galaxies: clusters: general}

\maketitle
\section{Introduction}

In the complex hierarchical network 
of galaxies, galaxy groups, clusters, and superclusters called 
the cosmic web,
the largest relatively isolated systems  are galaxy superclusters
\citep{1956VA......2.1584D, 1978MNRAS.185..357J, 
2015A&A...580A..69E, 2016A&A...588L...4L}.
This makes galaxy superclusters 
unique objects in the studies of formation
and evolution of the cosmic web at different redshifts.

Deep surveys 
make it possible to compile supercluster catalogues in wide redshift
intervals
\citep{2007A&A...462..397E, 2012A&A...539A..80L, 2014MNRAS.445.4073C}
or to determine individual superclusters  
at high redshifts \citep{2007MNRAS.379.1546T, 2007MNRAS.379.1343S,
2008ApJ...677L..89G, 2009AJ....137.4867L, 
2011A&A...532A..57S, 2011MNRAS.413..177G, 
2016A&A...592A...6P, 2016ApJ...821L..10K}.
Recently, \citet{2016A&A...588L...4L} reported the
discovery of a very massive extended supercluster complex at the redshift
$z = 0.47$ called the BOSS Great Wall (BGW), 
using the CMASS (constant mass) sample of the Sloan Digital Sky Survey III (SDSS-III) 
\citep{2011AJ....142...72E,  2013MNRAS.435.2764M, 2016MNRAS.455.1553R}. 
The superclusters in the BGW are bigger and richer than any other supercluster
at this redshift discovered at present \citep[see, e.g. ][]{2011A&A...532A..57S,
2016A&A...592A...6P}.
In this paper we focus on the study of the morphology of the BGW superclusters.
The analysis of supercluster morphology enables us to quantify their
outer shape and inner structure \citep{1997ApJ...482L...1S, 1998ApJ...508..551S,
bas03, 2003MNRAS.343...22S, sss04, 2007A&A...476..697E, 
2011A&A...532A...5E, 2011ApJ...736...51E, 2011MNRAS.411.1716C}.
The morphology of superclusters can be used
to compare observed and simulated superclusters and to test cosmological models
\citep{kbp02, 2007A&A...462..397E, 2007A&A...476..697E, 2011MNRAS.411.1716C, 
2013ApJ...777...74S, 2014arXiv1404.3639S}.
The supercluster environment affects the properties of galaxies in it
\citep{2007A&A...476..697E, e08, 
2009A&A...495...37T, 2011A&A...529A..53T, 2012A&A...545A.104L}. 
The morphology of superclusters is one environmental factor which
shapes the properties of galaxies and galaxy groups 
\citep{2014A&A...562A..87E}.
Several studies have shown that richer superclusters with
luminosities higher than approximately $4\times10^{12}h^{-2} L_{\sun}$ are larger
and more elongated than poor superclusters
\citep{2011A&A...532A...5E, 2011A&A...535A..36E, 2011MNRAS.411.1716C}.
\citet{2011A&A...535A..36E} employed the 
principal component analysis and
combined morphological and physical parameters of superclusters
to derive scaling relations for
supercluster luminosities.

The data concerning BGW superclusters gives us the opportunity for the first time
to compare the morphology of superclusters at different
redshifts. 
In this paper we used the BGW supercluster data to study the morphology
and to obtain the luminosities of these superclusters, using the scaling relations
derived from the principal component analysis.
We have used the 
relation between the stellar masses of the main galaxies in haloes
and halo mass to calculate supercluster masses,
find the mass-to-luminosity ratios of superclusters, and compare these
with those of the richest local superclusters.

In accordance with \citet{2016A&A...588L...4L}, and also with 
studies based on SDSS data, used for comparison 
\citep[for example, ][]{2011A&A...535A..36E, 2012A&A...539A..80L, 2016A&A...595A..70E}
we assumed  the standard cosmological parameters: the Hubble parameter $H_0=100~ 
h$ km~s$^{-1}$ Mpc$^{-1}$, the matter density $\Omega_{\rm m} = 0.27$, and the 
dark energy density $\Omega_{\Lambda} = 0.73$ \citep{2011ApJS..192...18K}.
    
\section{Data}

\begin{figure}%[ht]
\centering
\resizebox{0.43\textwidth}{!}{\includegraphics[angle=0]{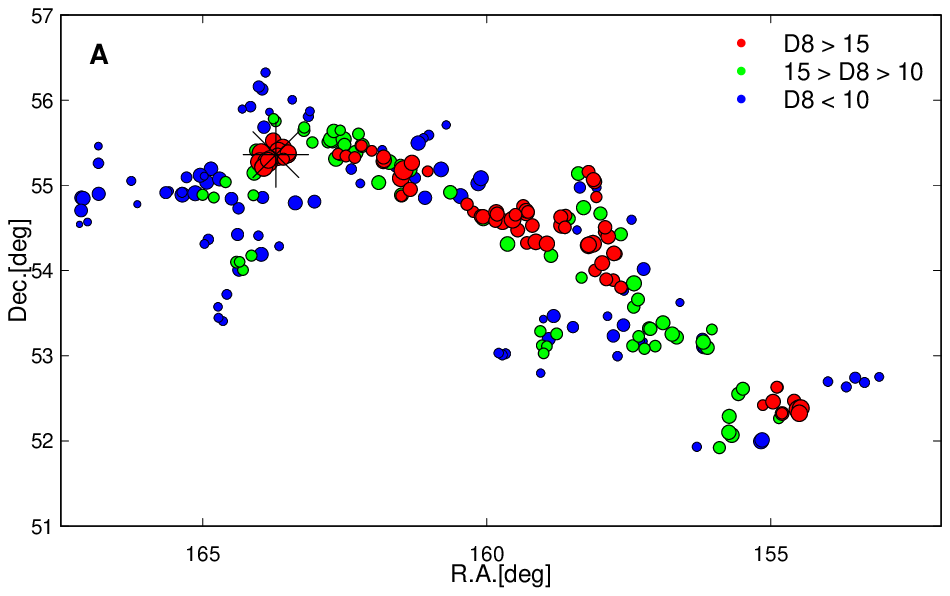}}\\
\resizebox{0.43\textwidth}{!}{\includegraphics[angle=0]{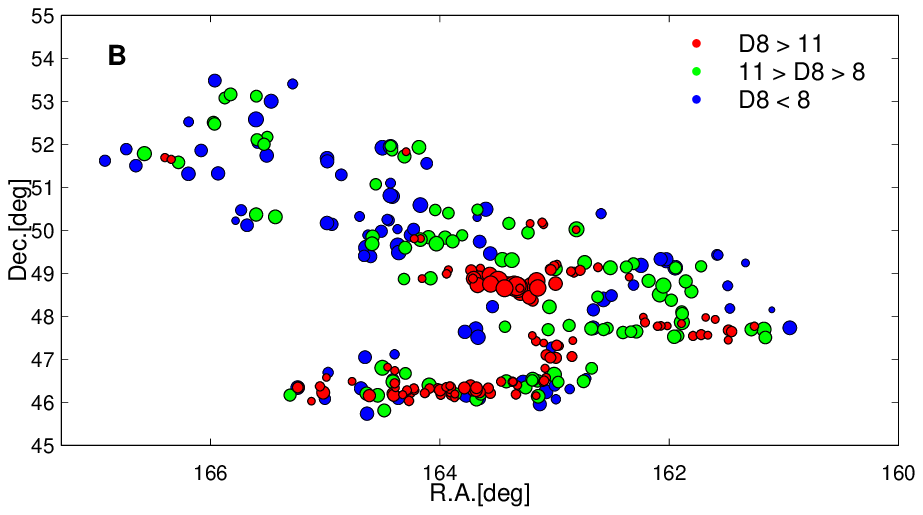}}\\
\resizebox{0.43\textwidth}{!}{\includegraphics[angle=0]{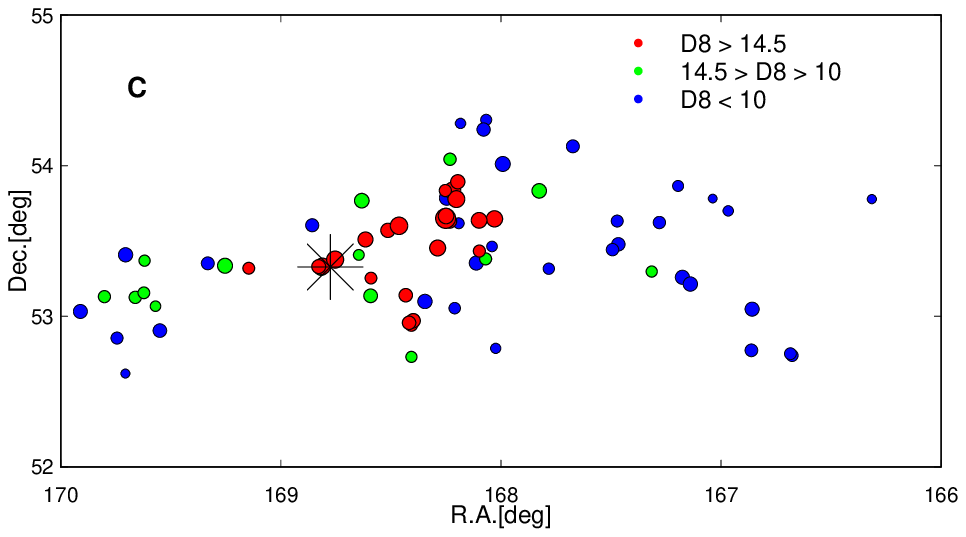}}\\
\resizebox{0.43\textwidth}{!}{\includegraphics[angle=0]{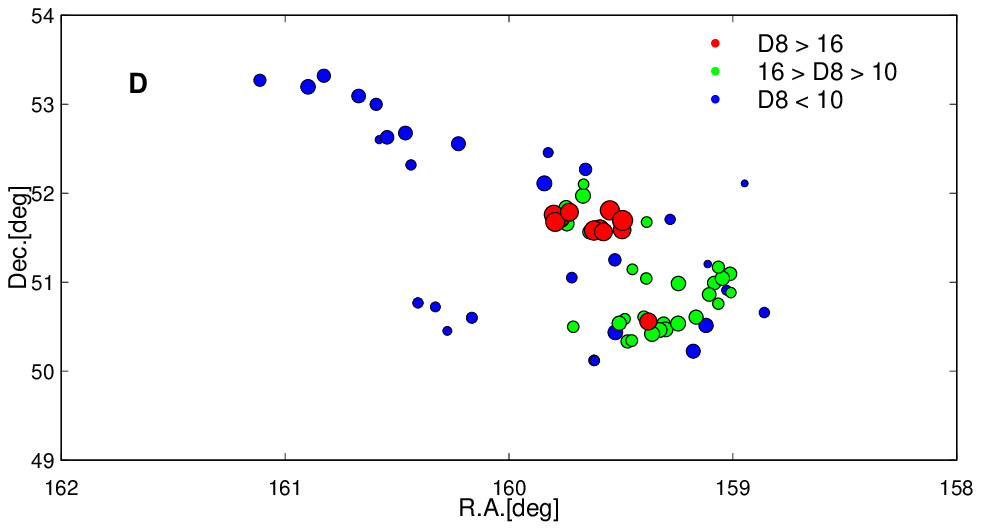}}
\caption{
Distribution of galaxies in the BGW superclusters in the sky plane (from up to down:
the superclusters A, B, C, and D). 
Red circles denote groups in the regions of the highest global
density in each supercluster; green circles correspond to groups with 
intermediate luminosity density, 
and blue circles correspond to groups with the lowest luminosity density,
as shown in the panels.
The density limits are chosen so that each density interval contains approximately
one-third of the supercluster galaxies.
Symbol sizes are proportional to the value of the density field at the
location of a galaxy.
The stars in the plots of superclusters A and C denote the location of the Planck
clusters PSZ2 G151.62+54.78 (supercluster A), 
and PSZ2 G150.56+58.32 (supercluster C).
}
\label{fig:radec}
\end{figure}

We used the data from the twelfth data release (DR12) of the 
SDSS \citep{2015ApJS..219...12A, 2000AJ....120.1579Y} 
Baryon Oscillation Spectroscopic Survey 
\citep[BOSS;][]{2011AJ....142...72E, 2012AJ....144..144B, 2013AJ....145...10D}. 
The BOSS data was published in the Data Release~8 \citep[DR8;][]{2011ApJS..193...29A}. 
From these data, we use
the CMASS (constant mass) sample, which selects massive and luminous galaxies 
in the redshift range $0.43<z<0.7$, with the stellar masses 
approximately constant up to $z\sim 0.6$ \citep{2013MNRAS.435.2764M}. 
This is the 
massive end of the red sequence. These are the most abundant galaxies 
at the high mass end ($M>10^{11}~M_{\odot}$) and they evolve passively over 
the CMASS redshift range \citep{2014arXiv1410.5854M, 2016MNRAS.456.3265M}.
For the details of our sample we refer the reader to \citet{2016A&A...588L...4L}.

Galaxy superclusters were determined using the
luminosity-density field 
following the same procedure that was used in \citet{2012A&A...539A..80L}. 
We weighted the luminosities of galaxies 
in the {\it r}-band  
to keep the mean density the 
same through the whole distance range, 
and then calculated the density field on a 3\,$h^{-1}$Mpc grid 
with a 8\,$h^{-1}$Mpc smoothing scale. 
The calculation of the luminosity-density field is described in 
\citet{2012A&A...539A..80L} and \citet{2014A&A...566A...1T}.

Superclusters of galaxies were defined as connected volumes
above a certain luminosity density threshold.  
\citet{2016A&A...588L...4L} analysed the properties of superclusters
at a series of density thresholds. They found an unusually high
overdensity at $D8 = 5$ level, in which at 
the density level $D8 = 6$  individual
superclusters can be distinguished from each other. 
Therefore, the BGW superclusters were extracted as connected volumes above the density
level $D8 = 6$ times the mean luminosity density 
($\ell_{\mathrm{mean}}$ = 
5$\cdot10^{-4}$ $\frac{10^{10} h^{-2} L_\odot}{(\vmh)^3}$)  
of the CMASS sample \citep{2016A&A...588L...4L}.

The BGW consists of two very rich superclusters with the diameters of 186\,$h^{-1}$Mpc 
(supercluster A) and 
173\,$h^{-1}$Mpc (supercluster B), 
and of two moderately large 
superclusters (superclusters C and D) 
with the diameters of 64 and 91\,$h^{-1}$Mpc. 
Data concerning the BGW superclusters are given in Table~\ref{BigScls}       
which presents the number of galaxies in superclusters,
the supercluster diameter (the maximum distance between the galaxies in
the supercluster), the supercluster volume (the number of connected  grid cells in the 
luminosity density field, multiplied by the cell volume), and 
the mean luminosity density in the supercluster, in  units of the mean luminosity density.
We show the sky distribution of galaxies in the BGW superclusters
in Fig.~\ref{fig:radec}. 
We plot galaxies in  the high, medium, and low luminosity density regions  
in superclusters with different colours. Each region contains approximately one third of the supercluster galaxies. 
The full configuration of the BGW superclusters
is shown in \citet{2016A&A...588L...4L}.

We also used the data from the Second Planck Catalogue of
Sunyaev-Zeldovich Sources (PSZ2) to identify Planck 
thermal Sunyaev-Zeldovich effect (tSZ) clusters in the
BGW region \citep{2015arXiv150201598P}.
Two Planck SZ sources 
were found: PSZ2 G150.56+58.32 in the region of the BGW supercluster
C, and PSZ2 G151.62+54.78 in the region of the supercluster A.
We show them in Fig.~\ref{fig:radec}. % and Fig.~\ref{fig:images}. 
These two Planck clusters correspond to the  
highest luminosity density of these individual superclusters.

\begin{table}
\caption{Data on the BGW superclusters.} 
\centering
\begin{tabular}{c r r r r }
\hline\hline
ID & Richness & $\mathrm{Diameter}$ & $\mathrm{Volume}$ & $D8_{\mathrm{mean}}$ \\
 & $N_{\mathrm{gal}}$ & $h^{-1}$Mpc & ($h^{-1}$Mpc)$^3$ & 
 \\
\hline       
A & 255 & 186.1 & 67500 & 9.1  \\
B & 303 & 172.9 & 70848 & 9.3  \\
C & 73  &  63.8 & 19008 &10.2  \\
D & 71  &  90.6 & 13635 & 9.3  \\
\hline
\end{tabular}
\label{BigScls}
\tablefoot{
Mean luminosity density in superclusters, $D8_{\mathrm{mean}}$,
is in units of the mean luminosity density calculated
with a 8\,$h^{-1}$Mpc smoothing scale (see text).
}
\end{table}    

\section{Methods}

\subsection{Minkowski functionals and shapefinders} 
\label{sect:mink} 

We employed the Minkowski functionals and shapefinders to study
the morphology of the BGW superclusters.
The BGW superclusters are defined as connected volumes above a certain luminosity 
density level, and can be characterised by their outer isodensity surface,
and their enclosed volume. 
The morphology and topology of the isodensity
contours are completely characterised
by four Minkowski functionals.
The Minkowski functionals were introduced in cosmology by 
\citet{1994A&A...288..697M}.
These functionals can be interpreted 
as the volume, the area, the integrated mean curvature 
(the first three Minkowski functionals), 
and the integrated Gaussian curvature (Euler characteristic)
of the isodensity surface (the fourth Minkowski functional)
\citep[see Appendix~\ref{sect:MF} and][for details and references]
{2007A&A...476..697E, 2011A&A...532A...5E, 2011ApJ...736...51E}. 
When increasing the isodensity level
over the threshold overdensity, we move into the central parts of
the supercluster. The Minkowski functionals can be calculated
for the full range of density levels from the full supercluster
to the central highest density peaks to show how the morphological
properties of a supercluster 
change with the increase of the isodensity level.

\citet{sah98} and \citet{sss04} used the first three Minkowski
functionals to calculate the shapefinders
$K_1$ (planarity) and $K_2$ (filamentarity),
 and their ratio, the shape parameter 
$K_1$/$K_2$ for the enclosed volume \citep[see also ][]{saar09}. 
The smaller the shape parameter, the more elongated a supercluster is.
The characteristic curve in the shapefinders $K_1$--$K_2$ plane 
is called the morphological signature \citep{2007A&A...476..697E}.
In the $K_1$/$K_2$-plane filaments are located near the $K_2$-axis and
pancakes are located near the $K_1$-axis.
Spheres are located at the origin of the plane where $K_1 = K_2 = 0$,  
and ribbons along the diagonal of the plane. 

The fourth Minkowski functional, $V_3$ (clumpiness),
characterises the inner structure of superclusters. It shows 
the number of isolated clumps, void bubbles, and tunnels in the enclosed
volume. When we increase the density level, 
the number of isolated clumps in a supercluster changes, void bubbles, and tunnels may
appear inside a supercluster, and this changes the value of $V_3$.
The higher the 
value of $V_3$, the more complicated is the inner morphology of the supercluster.

\subsection{Principal component analysis and luminosity of 
superclusters from scaling relation} 
\label{sect:pca} 

\citet{2011A&A...535A..36E} employed principal 
component analysis (PCA) to study the correlations 
between the physical and morphological properties of galaxy superclusters
drawn from the SDSS DR7 and to determine scaling relations for superclusters.
Principal component analysis finds a small number of 
linear combinations of correlated parameters  to describe most of the 
variation in the dataset with a small number of new uncorrelated parameters. PCA 
transforms the data to a new Cartesian coordinate system, where the greatest variance by 
any projection of the data lies along the first coordinate (the first principal 
component), the second greatest variance -- along the second coordinate, and so on. 
There are as many principal components as there are parameters, but typically 
only the first few are needed to explain most of the total variation. 
The principal components PC$x$  ($x \in \mathbb{N}$, $x \leq N_{\mathrm{tot}})$ 
are linear combinations of the original parameters:

\begin{equation}\label{eq:pc}
 PCx = \sum_{i=1}^{N_{\mathrm{tot}}} a_x(i) V_{i},
\end{equation}
where $-1 \leq a(i)_x \leq 1$ are the coefficients of the linear transformation,
$V_i$ are the original parameters and $N_{\mathrm{tot}}$ is the number of the original 
parameters.

\citet{1984MNRAS.206..453E} showed how to use PCA to get scaling relations. If, for example, the data points lie 
mostly along a plane, defined by the first two principal components, then the scaling 
relations for this plane are defined by the fact that the plane is perpendicular to the third principal component.
In general, if the data dimension is higher, 
the points may be concentrated around a hyperplane that is perpendicular 
to the principal component $PCy$ that we choose to ignore in the total variance:

\begin{equation}\label{eq:scaling}
 \sum_{i=1}^{N_{\mathrm{tot}}} a_y(i) \frac{(V_{i} - \overline{V_{i}})}{\sigma (V_{i})} = 0.
\end{equation}

\citet{2011A&A...535A..36E} studied the properties of galaxy superclusters
with principal component analysis and found that superclusters
can be characterised by a small number of physical and morphological
parameters, the diameter and shape parameters among them. 
They derived the scaling relations for superclusters
combining their morphological and physical parameters and
showed that luminous superclusters can be 
divided into more elongated and less elongated systems, with different scaling relations.
For elongated luminous superclusters the scaling relation 
from  \citet{2011A&A...535A..36E} is:
\begin{equation}\label{eq:scalingkl}
\log(L) = (0.22 K_2 - 1.67 K_1 + 1.45)\cdot\log(D) + 0.69,
\end{equation}
where $L$ is the total luminosity of the supercluster (in units of $10^{10}L_\odot$, $D$ is the supercluster
diameter (in units of $\mathrm{Mpc}/h$), and $K_1$ and $K_2$ are the shapefinders (planarity and filamentarity)
for the supercluster.
We found the morphological parameters for the
BGW superclusters and then  applied this scaling relation  
to calculate supercluster luminosities
in {\it r}-band,
denoted as $L^{\mathrm{scaling}}_{\mathrm{scl}}$.
We also calculated the masses of superclusters using their luminosities from 
scaling relation, $L^{\mathrm{scaling}}_{\mathrm{scl}}$, 
and the mass-to-luminosity ratio as found
for local rich superclusters  in the {\it r}-band 
\citep{2015A&A...580A..69E, 2016A&A...595A..70E}, $M/L \approx 300$~$h\,M_\odot$/$L_\odot$.
We denote the mass calculated in this way as $M^{\mathrm{scaling}}_{\mathrm{scl}}$.

\subsection{Masses and luminosities of the BGW superclusters from stellar
masses of galaxies}
\label{sect:masses} 

To estimate the minimum masses of the BGW superclusters
we adopted the same procedure as in \citet{2016A&A...588L...4L} who
used the stellar masses of galaxies 
to find the masses of the BGW superclusters.
The BOSS stellar masses are obtained from the Portsmouth galaxy product 
\citep{2013MNRAS.435.2764M}, which is based on the stellar population 
models by \citet{2005MNRAS.362..799M} and \citet{2009MNRAS.394L.107M}. The Portsmouth 
product uses an adaptation of the publicly-available Hyper-Z code 
\citep{2000A&A...363..476B} to perform a best-fit to the observed {\it{ugriz}} 
magnitudes of BOSS galaxies, with the spectroscopic redshift determined by the BOSS pipeline. 
The stellar masses used in this work were computed assuming 
the Kroupa initial mass function.

The virial masses of the host haloes of  galaxies
can be calculated from the  relation between the stellar masses of the
first ranked galaxies in haloes, $M_*$, and 
the virial masses of the haloes to which these galaxies belong, $M_{\mathrm{halo}}$
\citep{2010ApJ...710..903M}:
\begin{equation}\label{eq:mass}
\frac{M_{*}}{M_{\mathrm{halo}}}=2\left(\frac{M_{*}}{M_{\mathrm{halo}}}
\right)_0 \left[\left(\frac{M_{\mathrm{halo}}}{M_1}
\right)^{-\beta}+\left(\frac{M_{\mathrm{halo}}}{M_1}\right)^\gamma\right]^{-1},
\end{equation}
where $(M_*/M_{\mathrm{halo}})_0 = 0.0254$ is the normalization of the stellar to halo mass 
relation, $M_1=10^{11.95}$ is a characteristic mass, 
and $\beta=1.37$ and $\gamma=0.55$ are the slopes of the low 
and high mass ends of the relation, respectively \citep[][Table 6]{2010ApJ...710..903M}. 
The sum of the halo masses gives us an estimate of the lower limit
of the supercluster mass. 

In calculations of the masses of superclusters we
only used galaxies with stellar masses $\log(M_*/h^{-1}M_\odot) \geq 11.3$,
this is the completeness limit of the CMASS sample \citep{2013MNRAS.435.2764M}. 
We assumed that galaxies in the CMASS sample with $\log(M_*/h^{-1}M_\odot) \geq 11.3$
are the central galaxies of haloes.
This is based on comparison with 
the Sloan Digital Sky Survey main sample of galaxies as follows. 
We used the magnitude-limited friend-of-friends group catalogue from the SDSS DR10
main sample
by \citet{2014A&A...566A...1T} to select galaxies 
in superclusters with the luminosity density $D8 \geq 5$
in the distance bin from 180 to 270\,$h^{-1}$Mpc.  
From this sample we determined a BOSS CMASS-like high-mass sample of galaxies 
with a stellar mass limit of $\log(M_*/h^{-1}M_\odot) \geq 11.3$
and found that 87\,\% of all galaxies in this high-mass sample
are the most luminous (first-ranked) galaxies 
in the friend-of-friends groups, or they are single galaxies 
(the main galaxies of faint groups,
with satellite galaxies too faint to be 
observed in the SDSS). Therefore, we can assume that the high-mass galaxies 
in the BOSS sample are the first-ranked galaxies in groups.
Comparison
with  local galaxies suggests that this may introduce an error in mass estimates
of the order of about 10--15\,\%, considering that some massive galaxies may be
members of the same group and not the main galaxies of different groups.   

There are also haloes with the first-ranked galaxies having lower stellar
masses than the limit $\log(M_*/h^{-1}M_\odot) = 11.3$.
To take into account the mass in these haloes, we applied 
the scaling based on the analysis
of the Sloan Digital Sky Survey main sample of galaxies. 
We again used the data about galaxies in superclusters with the luminosity density $D8 \geq 5$
in the distance bin from 180 to 270\,$h^{-1}$Mpc, and found that the  
ratio between the total stellar mass in  high-mass galaxies 
and the total stellar mass in all first-ranked galaxies in the SDSS 
superclusters is 0.082 
($\approx 12$). 
The stellar masses of galaxies and halo masses are well correlated 
\citep{2010ApJ...710..903M}. Therefore we
used this ratio to scale our minimum 
supercluster mass estimates. 
Our mass estimates differ from what was adopted in \citet{2016A&A...588L...4L}
who used all galaxies in the BGW superclusters as the first-ranked galaxies
in calculations of halo masses. 
For details we refer to \citet{2016A&A...588L...4L}. 
We denote supercluster masses obtained from stellar masses of galaxies as
$M^{\mathrm{*}}_{\mathrm{scl}}$.

We also calculated the luminosities of superclusters as the sum of
the observed luminosities of high stellar mass galaxies 
in the {\it r}-band with a stellar mass limit 
of $\log(M_*/h^{-1}M_\odot) \geq 11.3$
as in calculations of masses of superclusters. To take into account
the luminosities of galaxies fainter than this limit we
used the ratio of the mean luminosity density in the CMASS sample
and the mean luminosity density of the SDSS MAIN galaxy sample
\citep[$\ell_{\mathrm{mean}}$ = 
1.65$\cdot10^{-2}$ $\frac{10^{10} h^{-2} L_\odot}{(\vmh)^3}$, see][]
{2012A&A...539A..80L}, corrected
for the mean overdensity in the BGW region.
These luminosities are denoted as $L^{\mathrm{lum}}_{\mathrm{scl}}$.

Below we calculate mass-to-light ratios 
$M^{\mathrm{*}}_{\mathrm{scl}}/L^{\mathrm{scaling}}_{\mathrm{scl}}$ for superclusters
as the ratio of the mass obtained from the stellar masses of galaxies,
and the luminosity of superclusters from the scaling relation.
As input, the scaling relation
uses morphological parameters of superclusters and supercluster 
diameters, being independent from other luminosity 
estimates that  use galaxy luminosities. 
We also compare the luminosities and masses of 
superclusters obtained with two different methods.

\section{Results}
\label{sect:results} 

\subsection{Morphology} 
\label{sect:morph}

\begin{figure}%[ht]
\centering
\resizebox{0.440\textwidth}{!}{\includegraphics[angle=0]{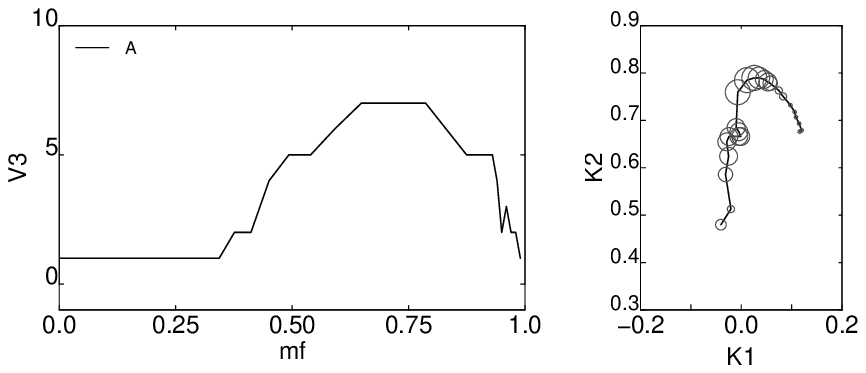}}
\resizebox{0.440\textwidth}{!}{\includegraphics[angle=0]{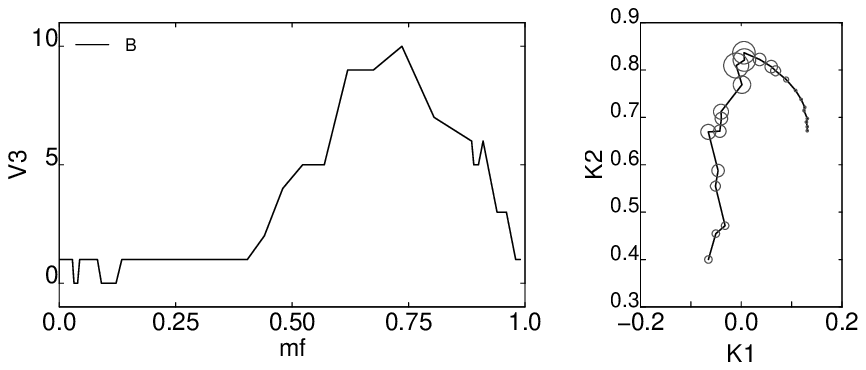}}
\resizebox{0.440\textwidth}{!}{\includegraphics[angle=0]{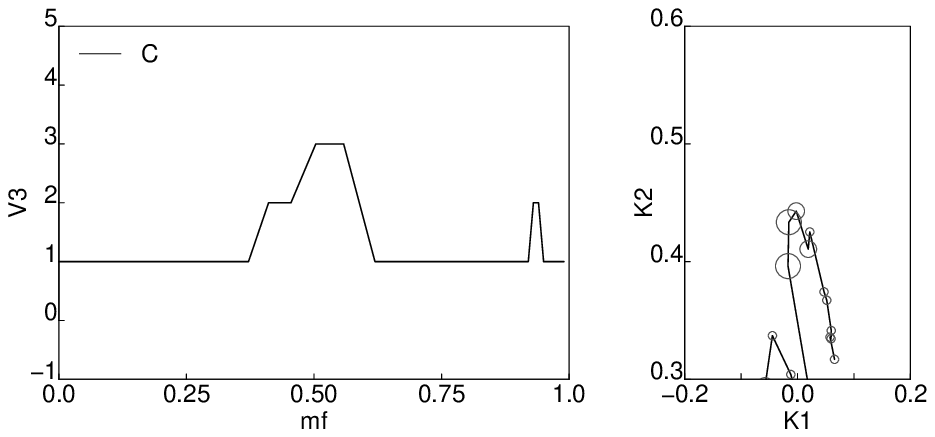}}
\resizebox{0.440\textwidth}{!}{\includegraphics[angle=0]{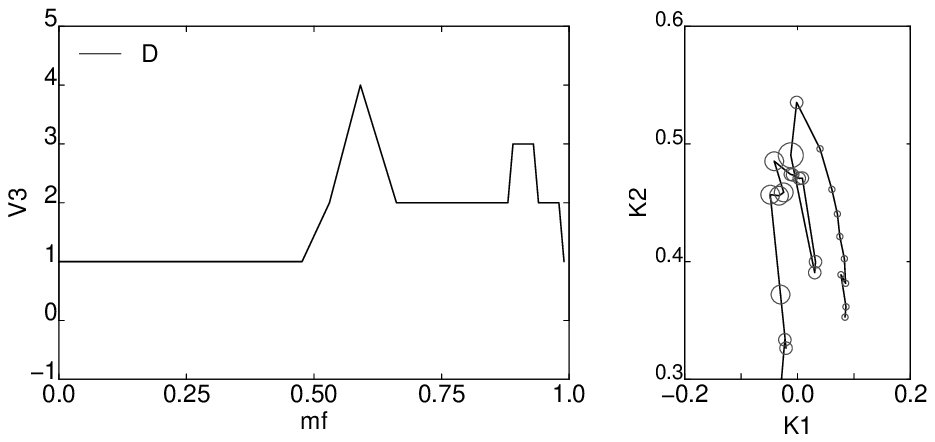}}
\caption{
Left panels: the fourth Minkowski functional $V_3$ versus the mass fraction $mf$
for the BGW superclusters.
Right panels: 
the shapefinders $K_1$ (planarity) and $K_2$ (filamentarity) plane
for a supercluster. The morphological signature in the $K_1-K_2$ plane 
is parametrically defined as $K_1(mf)$ and $K_2(mf)$. The right-hand
end of the $K_1$-$K_2$ curve corresponds to the whole 
supercluster (the mass fraction $mf = 0$); the mass fraction increases 
counterclockwise along the $K_1-K_2$ curve. 
The sizes of open circles are proportional 
to the value of $V_3$ at a given mass fraction $mf$. They show the change
of the clumpiness with the mass fraction together with the changes in the 
morphological signature. See text for more details.}
\label{fig:morph}
\end{figure}

\begin{table}
\caption{Morphological parameters for the BGW superclusters.} 
\centering
\begin{tabular}{rrrrrrr}
\hline\hline
ID & Richness & diameter  & $V_3$ & $K_1$ & $K_2$ & $K_1$/$K_2$  \\
 &  & $h^{-1}$Mpc & & & &  \\
\hline       
A & 255 & 186.1 &  7 & 0.12 & 0.68  & 0.17  \\
B & 303 & 172.9 & 10 & 0.13 & 0.67  & 0.19  \\
C & 73  &  63.8 &  4 & 0.07 & 0.32  & 0.21  \\
D & 71  &  90.6 &  3 & 0.08 & 0.35  & 0.24  \\
\hline
\end{tabular}
\label{tab:morphl}
\tablefoot{
Columns are as follows:
(1) Notation (ID); 
(2) the number of galaxies;
(3) the diameter;
(4) the maximum value of the fourth Minkowski functional $V_3$ (clumpiness);
(5) the value of the shapefinder $K_1$ (planarity);
(6) the value of the shapefinder $K_2$ (filamentarity);
(7) the value of the shape parameter $K_1$/$K_2$.
}
\end{table}    

\begin{table*}
\caption{Luminosities, masses, and mass-to-light ratios of the BGW superclusters.}
\centering
\begin{tabular}{rrrrrr}
\hline\hline
ID & $L^{\mathrm{lum}}_{\mathrm{scl}}$ & $L^{\mathrm{scaling}}_{\mathrm{scl}}$ 
& $M^{\mathrm{*}}_{\mathrm{scl}}$  
& $M^{\mathrm{*}}_{\mathrm{scl}}/L^{\mathrm{scaling}}_{\mathrm{scl}}$ & $M^{\mathrm{scaling}}_{\mathrm{scl}}$   \\
 & $10^{12}h^{-2} L_{\sun}$ & $10^{12}h^{-2} L_{\sun}$  
 & $10^{16}h^{-1}M_\odot$ & $h\,M_\odot$/$L_\odot$  & $10^{16}h^{-1}M_\odot$ \\
\hline       
A & 82.1  & 75.9  & 2.1  & 277 & 2.3 \\
B & 47.9  & 60.3  & 1.1  & 182 & 1.8 \\
C & 28.7  & 17.2  & 0.7  & 407 & 0.6 \\
D & 10.9  & 25.4  & 0.3  & 118 & 0.8 \\
BGW&169.6 &178.4  & 4.2  & 234 & 5.3 \\
\hline
\end{tabular}
\label{tab:masses}
\tablefoot{
Columns are as follows:
(1) Notation; 
(2) the luminosity, calculated using the luminosities of galaxies with 
stellar masses  $\log(M_*/h^{-1}M_\odot)\geq 11.3$, corrected for faint galaxies
as described in the text;
(3) the luminosity, calculated using the scaling relation;
(4) the mass  as sum of halo masses of galaxies
with stellar masses  $\log(M_*/h^{-1}M_\odot)\geq 11.3$, corrected for faint galaxies;
(5) the mass-to-light ratio, $M^{\mathrm{*}}_{\mathrm{scl}}/L^{\mathrm{scaling}}_{\mathrm{scl}}$;
(6) the mass obtained from the scaling relation (see text).
Luminosities are in the {\it r}-band.} 
\end{table*}    

In Fig.~\ref{fig:morph} we present the fourth Minkowski functionals $V_3$ 
and the morphological signatures for the BGW superclusters.
The values of the morphological parameters are given in Table~\ref{tab:morphl}.
For the argument labelling the isodensity surfaces
in Fig.~\ref{fig:morph} (left panels) 
we use the (excluded)  mass 
fraction $mf$ -- the ratio of the mass in the regions with lower density than 
at the isodensity surface, to the total mass of the supercluster. 
For the whole supercluster $mf=0$, this corresponds to the lowest value
of the threshold density used to determine the supercluster. 
The mass fraction $mf=1$ corresponds to the  peak density
in the supercluster high-density cores.
The fourth Minkowski functional $V_3$ is calculated for a full
range of mass fractions from $mf=0$ to $mf=1$.

Figure~\ref{fig:morph} shows that at the mass fraction $mf \approx 0.3$,
where approximately one third of galaxies from the outskirts of the supercluster
do not contribute to the supercluster, the value of $V_3$
starts to increase. These outskirts regions are plotted in
Fig.~\ref{fig:radec}. 
At higher values of $mf$ superclusters
become clumpy, they may split into several high-density cores, 
and may have void bubbles or tunnels in them.
At a certain $mf$ level the value of $V_3$ reaches maximum.
For rich BGW superclusters
this happens at approximately $mf = 0.7$ (Fig.~\ref{fig:morph}), similarly   
to the local rich superclusters \citep{2007A&A...476..697E, 2011A&A...532A...5E}.
This mass fraction approximately marks the crossover from the lower density outskirts
of the superclusters to the high-density cores. 
The galaxies in the high-density cores of the BGW superclusters are plotted in Fig.~\ref{fig:radec}, where we give also the density levels which approximately correspond
to $mf = 0.7$. 
The BGW superclusters A and B have higher values of $V_3$ ($7$ and $10$), showing that they have a more
complicated and richer inner structure than the  BGW superclusters C and D
(with $V_3 = 4$ and $3$).
The clumpiness of the largest BGW supercluster (A) is lower 
than the clumpiness of the second largest supercluster. This may be due
to a very high density core region of the A supercluster, which has a 
smaller number of individual clumps than the core region of the B supercluster.
For poor BGW superclusters the maximum of $V_3$ occurs at a slightly lower $mf$ level. 
At still higher density levels less and less galaxies contribute to the
superclusters and the value of the clumpiness $V_3$ decreases.
High peaks in the $V_3$ distribution at high mass fractions for
poor BGW superclusters suggest
that they have high-density compact clumps in core regions. 
In the supercluster C one of these clumps corresponds to the Planck cluster
PSZ2 G150.56+58.32.
At high density levels the supercluster D splits into two parts,
and the $V_3$ value decreases to two and increases again at very high density levels.

The richest local superclusters in the Sloan Great Wall (SGW) 
have maximum values of the clumpiness $V_3 = 13$ 
(the richest SGW supercluster) and $6$ (the second richest SGW supercluster),
close to the $V_3$ values of the rich BGW superclusters
\citep{2011A&A...532A...5E}.
These superclusters contain several high-density cores, as the high $V_3$ values
suggest also for the rich BGW superclusters \citep{2016A&A...595A..70E}.
The BGW supercluster C can be compared with the local A2142 supercluster which
also has one very rich galaxy cluster in its main body,
but morphologically the A2142 is different, it consists of one rich straight chain of galaxy
groups and clusters \citep{2015A&A...580A..69E}.

The right panels of Fig.~\ref{fig:morph} show how the morphological
signatures of superclusters change with the mass fraction. 
The planarity $K_1$ has its maximum value at $mf=0$. When the mass fraction
increases, the planarity  $K_1$ of a supercluster decreases and 
the  filamentarity  $K_2$ increases (counterclockwise from right 
to left in the right panels of Fig.~\ref{fig:morph}).
The changes in the filamentarity $K_2$ at low mass fractions are more rapid
than the changes in the planarity $K_1$, 
showing that superclusters become
more elongated. 
At the mass fraction of about $mf = 0.7$, 
the filamentarity $K_2$ reaches a maximum and then decreases showing that
in the high-density cores where approximately one-third of supercluster galaxies reside,
the morphology of a supercluster changes. The 
filamentarity  $K_2$ decreases rapidly, and the planarity
$K_1$ also decreases slightly. 
In this respect the BGW superclusters are similar to the richest
local superclusters
\citep{2007A&A...476..697E, 2011A&A...532A...5E}. 
At high density levels at which the supercluster D divides into two,
the morphological signature changes rapidly, too.

The largest difference between the local superclusters and the BGW superclusters
is their overall shape - the BGW superclusters are very elongated, having 
lower shape parameter values for the whole superclusters 
than any local supercluster studied so far 
\citep[$ < 0.25$ versus $> 0.25$ for local
rich superclusters, ][]{2011A&A...532A...5E}.

\subsection{The luminosities of superclusters} 
\label{sect:pca} 

In  Table~\ref{tab:masses} we show the luminosities
of the BGW superclusters obtained using the luminosities of
high stellar mass galaxies, as described in Sect.~\ref{sect:masses}. For comparison, we 
also present the luminosities obtained by the scaling relation
(Eq.~(\ref{eq:scalingkl})). This relation uses the data about 
supercluster diameters and
shapefinders (Tables~\ref{BigScls}-\ref{tab:masses}).
Table~\ref{tab:masses} shows that the values of luminosities from two methods
agree well, within 20\% for rich BGW superclusters, and up to 2.7 times
for poor BGW superclusters.

We can compare the luminosities of the BGW superclusters
with the luminosities of superclusters from the SDSS by \citet{2012A&A...539A..80L},
\citep[see also ][]{2012A&A...542A..36E, 2016A&A...595A..70E}.
The most luminous local superclusters
are the richest SGW superclusters, with the luminosities 
$51.6\times10^{12}h^{-2} L_{\sun}$ for the richest supercluster,
and $29.2\times10^{12}h^{-2} L_{\sun}$ for the second richest supercluster.
Table~\ref{tab:masses} shows that the luminosity of the BGW A supercluster
is almost equal to the sum of the luminosities of the two richest
SGW superclusters. The BGW A supercluster
is as large as these superclusters together therefore 
this result suggests that our methods to calculate the 
BGW supercluster luminosities work reasonably well.

\subsection{The masses and mass-to-light ratios of the BGW superclusters}
\label{sect:sclmass} 

To find supercluster masses from the stellar masses of galaxies
in superclusters we
used the relation between stellar masses
of galaxies and halo masses, as described in  Sect.~\ref{sect:masses}. 
The sum of halo masses was corrected for masses
of missing haloes. 
The masses of superclusters, $M^{\mathrm{*}}_{\mathrm{scl}}$,
are given in Table~\ref{tab:masses}.

The mass
of the second largest supercluster (B) in the BGW is approximatey
half of the mass of the largest supercluster (A). This is because
in this supercluster there are less very high stellar mass galaxies 
than in the supercluster A. Also \citet{2016A&A...588L...4L}
noted that the stellar masses of the superclusters A and B
are different with a high significance.
The highest mass halo in the supercluster C can be identified with the
Planck cluster PSZ2 G150.56+58.32  with the mass
of $M \approx 7.6\times~10^{14}h^{-1}M_\odot$ \citep{2015arXiv150201598P}.
This makes the mass
of this supercluster more than two times higher than the mass of another
poor BGW supercluster D. 
Among Planck clusters in the BGW redshift range, $0.47 \pm 0.05$,
this cluster has the highest estimated mass. 
The mass of the lowest mass BGW supercluster D
is similar to the mass of another supercluster at redshift $z \approx 0.5$,
the SCL2243-0935 supercluster \citep{2011A&A...532A..57S}.
The second  
Planck cluster, PSZ2 G151.62+54.78 in the supercluster A,
is less massive with $M \approx 5.4\times~10^{14}h^{-1}M_\odot$.

In Table~\ref{tab:masses} we also present supercluster masses obtained
from their luminosities, estimated by the scaling relation.
These two mass estimates coincide best for the BGW
superclusters A and C.  For superclusters B and D the difference between
the two mass estimates is larger. 

\citet{2016A&A...595A..70E} estimated the masses of the SGW superclusters
using several methods. 
The richest SGW supercluster has the total  mass of about 
$1.2 - 1.4 \times~10^{16}h^{-1}M_\odot$. The mass range of other 
SGW superclusters and some other local superclusters
is of about $0.3 - 0.7\times~10^{16}h^{-1}M_\odot$
\citep{2015A&A...580A..69E, 2016A&A...595A..70E}. 
Thus the BGW supercluster
A has a higher mass than very rich superclusters in the richest local
galaxy system. The masses of other BGW superclusters
are in the same range as the masses of rich local superclusters.
We note that superclusters contain also intracluster
gas. Therefore the BGW superclusters masses obtained by us in this
work are the lower mass limits only.

We also present in Table~\ref{tab:masses} the mass-to-light ratios 
$M^{\mathrm{*}}_{\mathrm{scl}}/L^{\mathrm{scaling}}_{\mathrm{scl}}$
for the BGW superclusters, calculated using the 
luminosity of superclusters as obtained from the scaling relation 
(Eq.~\ref{eq:scalingkl}), and mass from stellar masses of galaxies.
The values of the $M^{\mathrm{*}}_{\mathrm{scl}}/L^{\mathrm{scaling}}_{\mathrm{scl}}$ 
ratios for the largest BGW supercluster, 
$M/L = 277$ and $182$~$h\,M_\odot$/$L_\odot$, are close to the values of the mass-to-light ratios
of the largest two SGW superclusters, determined in the {\it r}-band, 
$M/L = 271$ and $241$~$h\,M_\odot$/$L_\odot$ \citep{2016A&A...595A..70E}.
The $M^{\mathrm{*}}_{\mathrm{scl}}/L^{\mathrm{scaling}}_{\mathrm{scl}}$
ratio of the whole BGW is close to the value of the 
$M/L$
of the whole SGW, $234$ and $272$, correspondingly.
The supercluster B has 
$M^{\mathrm{*}}_{\mathrm{scl}}/L^{\mathrm{scaling}}_{\mathrm{scl}} \approx 200$~$h\,M_\odot$/$L_\odot$. The same value of the mass-to-light
ratio was obtained for the A901/902 supercluster at redshift $z = 0.165$
in the {\it r}-band \citep{2008MNRAS.385.1431H}.
For the poor supercluster D we obtained the mass-to-light ratio 
$118$~$h\,M_\odot$/$L_\odot$. 
This is similar to the $M/L$ of the galaxy filaments 
in the SCL2243-0935 supercluster 
\citep[][in this paper the luminosities have been 
found in the {\it i}-band]{2011A&A...532A..57S}.

\subsection{The uncertainties of morphology,
luminosities and masses of the BGW superclusters} 
\label{sect:unc}

{\it Morphological parameters}.
Uncertainties in calculation of the morphological parameters of 
superclusters come
from how precicely we can calculate their values at low mass fractions,
and from the choice of the density level used to define superclusters.
We estimated that the uncertainties of the shapefinders $K_1$ and $K_2$ 
at low mass fractions are of the order of less than 5\%. 
The superclusters were determined at fixed luminosity density
level. If we decrease the density level used to define superclusters, 
new galaxies may be added to superclusters. Therefore, new clumps may
appear at the outskirts of superclusters which may change the value
of the fourth Minkowski functional $V_3$ at low mass fractions.
The possible change in the morphological parameters is individual
for each supercluster as discussed also in \citet{2011A&A...532A...5E}.
The BGW superclusters form a complex with overall very high
luminosity density. The individual BGW
superclusters were determined at the luminosity density level
$D8 = 6$. Already at the density level $D8 = 5.5$ superclusters A and C, and superclusters
B and D join to form two systems, and  at $D8 = 5$ they join into one huge system 
\citep[see][ for details]{2016A&A...588L...4L}.
These systems have a clumpiness and overall shape which differ from those
of individual superclusters, and cannot be compared with
the MFs of individual superclusters. 
\citet{2011ApJ...736...51E} showed that at lower density levels
at which the SGW superclusters join into huge systems
the clumpiness and the shape parameter values  increase, and that joint
systems are more planar than the individual SGW superclusters.

{\it Luminosity of superclusters}. 
The errors in the luminosity
obtained with Eq.~(\ref{eq:scalingkl}) are related to how precicely we
can calculate the values of the shapefinders $K_1$ and $K_2$.
As we showed above these uncertainties are of the order of less than 5\%.
These errors cause approximately 1\% uncertainty in luminosity,
calculated using Eq.~(\ref{eq:scalingkl}). 
This agrees with \citet{2011A&A...535A..36E}, who showed that the PCA results for superclusters
depend only very weakly on the choice of the density level used to define the
superclusters.
Another source of uncertainty comes from diameter errors.
The diameter of a supercluster is defined as the maximum distance between
supercluster galaxies. The main uncertainty of diameters is related
to how robust is the supercluster definition using a fixed luminosity density
level. If we decrease the density level, 
the superclusters A and C, and superclusters
B and D join to form two systems, but no new galaxies join superclusters
A, C, and D at their farthest edges where they could increase the diameter. 
The supercluster B has two galaxies added in one edge, the maximum increase
in diameter is approximately $8$~\Mpc\ which leads to the 5\% uncertainty
in the luminosity of the supercluster. 

The main source of uncertainty in luminosity calculations
using the luminosity of massive galaxies 
comes from unobserved galaxies
in the BGW superclusters. 
For the  BGW A supercluster, our luminosity estimates have closer values than for
the BGW B supercluster.
It is possible that the galaxy content of the BGW superclusters
A and B is different, with B containing relatively more faint galaxies.
\citet{2016A&A...588L...4L} showed that the distribution of galaxy stellar masses
in these superclusters is different, hinting that
their galaxy content differ. 
The luminosities obtained by the two methods for poor superclusters differ 
up to 2.5 times, showing that the
uncertainties are higher for poorer superclusters.
We emphasise that when estimating supercluster
luminosities by the scaling relation we used data
about the diameter and shape parameters of superclusters.
Even so, the values of luminosities obtained with two different methods
are in good agreement, suggesting that uncertainties from
unobserved galaxies in calculations of supercluster luminosities using galaxy
luminosities, $L^{\mathrm{lum}}_{\mathrm{scl}}$, were taken into account correctly.

{\it Masses of superclusters}. 
The errors of supercluster masses were estimated using the stellar 
mass error estimates from \citet{2013MNRAS.435.2764M}
who found that the average errors of $\log(M_*/M_\odot)$ are 0.1 dex.
We recalculated the supercluster masses 1000 times using 
up to 0.1 dex random deviations of galaxy stellar mass values 
(assuming the Gaussian distribution of errors) 
in the calculations of halo masses. This gives $1\sigma$ errors of masses
for the two richer superclusters as  0.05 and 0.06 dex, and
for the superclusters C and D  as 0.1 dex. After correcting for faint
galaxies (Sect.~\ref{sect:masses}, 
approximately 12 times) we find that
the uncertainties in supercluster total mass estimates are of the order 
up to 15\%.

\section{Discussion and conclusions}
\label{sect:disc}

{\it Supercluster shape parameters.}
Our study of the morphology of the BGW superclusters
showed that they are very elongated systems. The
overall morphology of the BGW superclusters is similar to the
morphology of local superclusters, with the maximum values of the
fourth Minkowski functional $V_3$ up to 10.  However, the BGW superclusters
are more elongated, with the lower value of the shape parameter
than local superclusters, $K1/K2 < 0.2$. 

\citet{2007A&A...462..397E} 
found that the richest 
superclusters from observations and simulations
have the shape parameters $K1/K2 > 0.2$, similar results were obtained
in \citet{2011MNRAS.411.1716C}. 
We use the CMASS luminous red galaxy data to trace
the BGW superclusters, and 
owing to the morphology-density relation
one might expect that luminous red galaxies trace the inner parts 
of superclusters only, making them to appear more elongated than they really
are. For the local superclusters we can compare the Minkowski functionals
and shapefinders for different galaxy populations \citep{e08}.
This comparison  showed that the Minkowski functionals
and shapefinders for bright and faint, and for elliptical and spiral
galaxies for the richest local superclusters in the outskirts of superclusters
have similar values. Therefore we cannot conclude that
very elongated shapes of the BGW superclusters are related to the
use of the data for red galaxies only. 

Another possible reason for the difference between the morphology of the BGW
superclusters and the richest local superclusters may be related to the 
evolution of supercluster morphology.
\citet{2009MNRAS.399...97A} analysed the future evolution
of galaxy superclusters in an acceleratingly expanding Universe
and showed that in the future, superclusters will separate from each other
and become less elongated. However, the redshift difference
between the local supercluster sample and the BGW superclusters
is not large ( 0.1 and 0.5) and it is questionable whether
we should expect strong morphology evolution in this interval. 
To understand this, a study of supercluster  morphology
in a wide redshift interval is needed.

{\it Supercluster masses, luminosities and mass-to-light
ratios.}
We determined the luminosities and masses of the BGW superclusters 
with two methods, using the luminosities of high stellar mass galaxies 
in superclusters, corrected for the faint galaxies missing from the 
survey, and employing the morphological parameters filamentarity and planarity, and the
diameters of superclusters. 
The values of luminosities and masses agree within approximately 20\% for rich
BGW superclusters, and up to 2.7 times for poor BGW superclusters
suggesting that uncertainties in luminosity and mass calculations were taken into
account properly.

We may expect
that the BGW superclusters have higher luminosities than local
superclusters of similar richness: the
luminosity evolution of stellar populations of galaxies 
implies  a $\approx 0.5-0.7$ mag dimming in the $r$-band, 
depending on the absolute age assumed as $z \approx 0.5$
\citep{2005MNRAS.362..799M, 2016MNRAS.456.3265M}.
However, we need to study a larger sample of superclusters to understand
the evolutionary trends of supercluster properties.

We compare the masses of the richest BGW superclusters 
with the masses of the superclusters from the SGW below.
The mass of the BGW supercluster C is dominated by a Planck cluster
with the mass of $M \approx 7.6\times~10^{14}h^{-1}M_\odot$;
this is approximately one-tenth of the total mass of the supercluster
according to our mass estimate. For comparison we
mention that in the supercluster A2142 
approximately one-fifth of the supercluster mass comes
from the mass of a very rich galaxy cluster A2142 
\citep{2015A&A...580A..69E}.
The values of the mass-to-luminosity ratios of the BGW superclusters
are comparable 
to these of the local rich superclusters.

{\it The BGW and SGW.}
The richest local supercluster complex is the SGW which has a 
total diameter of $230$~\Mpc, smaller than the BGW. 
The SGW consists of two rich and three poor superclusters
\citep{2016A&A...595A..70E}. 
These superclusters were defined as connected overdensity regions in the
luminosity density field at the density level 5 (in units of the mean luminosity density).
At a slightly lower density level, $4.7$, the superclusters from the SGW join into
one huge system \citep{2011ApJ...736...51E}. This is similar to the BGW supercluster
complex which forms one huge system at the luminosity density level 5 \citep{2016A&A...588L...4L}. 

We showed that the clumpiness of the largest SGW supercluster is even higher
than the clumpiness $V_3$ of the BGW supercluster A (13 vs. 10),
which shows that this supercluster has a more complicated inner structure than
the BGW A supercluster. The second richest superclusters in the BGW and the SGW
have close $V_3$ values (7 and 6). 
The largest difference between the morphology of the BGW and SGW superclusters
is that the BGW superclusters are very elongated, having 
shape parameter values $K_1 /K_2 < 0.20$ while for the richest SGW superclusters
 $K_1 /K_2 = 0.28$ and $0.48$ \citep{2011A&A...532A...5E}. 

The comparison of the masses of the BGW superclusters
with the masses of superclusters from the SGW shows that the most
massive BGW supercluster (A) has a higher mass than any
SGW supercluster  \citep[see
][for the masses of the SGW superclusters]{2016A&A...595A..70E}.
\citet{2016A&A...595A..70E} estimated that the 
lower mass limit of the SGW is $M = 2.4\times~10^{16}h^{-1}M_\odot$,
which is comparable to the mass of the BGW supercluster A,
$M \approx 2.1\times~10^{16}h^{-1}M_\odot$, 
and approximately two times 
lower than  the mass of the whole BGW, $M = 4.2\times~10^{16}h^{-1}M_\odot$. 

Similarly, the most luminous BGW supercluster (A) has the luminosity comparable to the
sum of luminosities of  the richest two
SGW superclusters. In addition, \citet{2016A&A...595A..70E}
estimated that the total luminosity of the SGW is approximately
$0.9\times10^{14}h^{-2} L_{\sun}$ while in this paper we
obtained that the total luminosity of the BGW is 
twice as high, $1.8\times10^{14}h^{-2} L_{\sun}$.
This shows that the BGW supercluster complex  
is richer, more massive and larger than the richest supercluster
complex in the local Universe, the SGW.

{\it Supercluster complexes.}
\citet{2011A&A...532A...5E} used data on the galaxy superclusters derived from
the SDSS MAIN data to describe the morphology and large-scale distribution
 of superclusters in the local Universe. They showed that at the distance interval of $210 - 260$~\Mpc\ 
superclusters form three chains, separated by voids. One chain is 
formed by the SGW, other chains are formed by 
the Bootes, the Ursa Major, and other rich superclusters. All of these superclusters
are poorer than the superclusters in the BGW. 
The very rich Corona Borealis supercluster is located at the intersection of
supercluster chains, and it is separated from the SGW by voids. Therefore 
we cannot consider this supercluster as a member of a common complex
with the SGW. There are also other very rich superclusters 
in the local Universe, like the Shapley, 
the Horologium-Reticulum, the Sculptor and others, 
but they are not located close to each other and  do not 
form supercluster
complexes like the BGW or the SGW \citep{1997A&AS..123..119E, 2005AJ....130..957F, 2006A&A...447..133P, 2011MNRAS.415..964L}. 

At redshifts of $0.5$ and higher, just a few superclusters have been found so far
\citep{2007MNRAS.379.1546T, 2007MNRAS.379.1343S,
2008ApJ...677L..89G, 2011A&A...532A..57S, 
2016A&A...592A...6P}. \citet{2016ApJ...821L..10K} mention that 
a supercluster at the redshift $z = 0.9$ may be embedded in a $\approx 100$~\Mpc\
overdense structure but no rich supercluster complexes have yet been found  
at so high redshifts.

{\it Supercluster morphology: filaments and spiders.}
Rich superclusters show wide morphological variety in 
which \citet{2011A&A...532A...5E} determined two main morphological types: 
spiders and filaments.  
Filament-type superclusters 
have elongated main bodies that connect galaxy groups and clusters in superclusters. 
Spider-type superclusters are systems of one or several high-density clumps 
with a large number of outgoing filaments connecting them.  
Empirical models show that morphological signatures of rich
superclusters correspond to the multibranching systems of filament and spider types;
poor superclusters typically are of spider type with one rich cluster
surrounded by outgoing galaxy chains \citep{2007A&A...476..697E}. 
\citet{2011A&A...532A...5E} classified superclusters 
 as filaments and spiders on the basis of their morphological
information and visual appearance.
The BGW superclusters can be classified as being of filament-type
(perhaps the supercluster C with its rich cluster and outgoing galaxy chains
can be classified as a spider-type supercluster). 
Simulations show that while the sizes of the richest observed and 
simulated superclusters are comparable \citep{2012ApJ...759L...7P},
the morphological variety of the observed superclusters is not 
recovered in simulations yet \citep{2007A&A...476..697E}.
In particular, very dense and large filament-type superclusters
like the richest SGW and BGW superclusters 
were not found among the richest simulated superclusters
\citep{2007A&A...476..697E}.
This shows the need to study the richest superclusters and their complexes
 from observations and simulations in combination with the analysis of 
supercluster morphologies.

\citet{2011MNRAS.417.2938S} 
applied extreme value statistics to show that the 
presence of such a massive and dense structure as the SGW is difficult 
to reconcile with the assumption of Gaussian
initial conditions if $\sigma_8$ is less than 0.9. 
They mention that this tension can be reduced if this structure 
is the densest within the Hubble volume. However, the discovery of the BGW 
shows that even richer systems exist. 
In addition, \citet{2006A&A...459L...1E} showed that 
the fraction of very luminous superclusters among the observed superclusters 
is higher than among simulated superclusters. 
The study of the properties and evolution
of very rich superclusters and their complexes 
using simulations with very large volume
may constrain the assumptions on initial conditions.

In summary, the analysis of the morphology and luminosity of the
BGW superclusters shows that the BGW 
is a unique complex of very rich and
luminous superclusters having a very elongated shape and a complicated inner 
structure. 
The search for and detailed study of the richest superclusters and their complexes
from observations and simulations in combination with the analysis of 
supercluster morphologies help us to understand the 
properties of the cosmic web and to constrain initial conditions.

\section*{Acknowledgements}

We thank the referee for comments that helped to improve the paper.
ET, LJL, ME, MG, and ES were supported by institutional research funding IUT26-2 and IUT40-2  of
the Estonian Ministry of Education and Research, and by the
Centre of Excellence “Dark side of the Universe” (TK133) financed by the
European Union through the European Regional Development Fund.
HL is supported by Turku University Foundation. 
AS, and JAR-M  acknowledge financial support from the Spanish Ministry
of Economy and Competitiveness (MINECO) under the 2011 Severo
Ochoa Program MINECO SEV-2011-0187.
AMD acknowledges support from the U.S. Department of Energy, Office of Science, Office of High Energy Physics, under Award Number DE-SC0010331. AMD also thanks the Center for High Performance Computing at the University of Utah for its support and resources.

In this work we used the R statistical environment
\citep{ig96}.

Funding for SDSS-III has been provided by the Alfred P. Sloan Foundation, 
the Participating Institutions, the National Science Foundation, 
and the U.S. Department of Energy Office of Science. 
The SDSS-III web site is http://www.sdss3.org/.

SDSS-III is managed by the Astrophysical Research Consortium 
for the Participating Institutions of the SDSS-III Collaboration 
including the University of Arizona, the Brazilian Participation Group, 
Brookhaven National Laboratory, Carnegie Mellon University, 
University of Florida, the French Participation Group, 
the German Participation Group, Harvard University, 
the Instituto de Astrofisica de Canarias, 
the Michigan State/Notre Dame/JINA Participation Group, 
Johns Hopkins University, Lawrence Berkeley National Laboratory, 
Max Planck Institute for Astrophysics, 
Max Planck Institute for Extraterrestrial Physics, 
New Mexico State University, New York University, Ohio State University, 
Pennsylvania State University, University of Portsmouth, 
Princeton University, the Spanish Participation Group, 
University of Tokyo, University of Utah, Vanderbilt University, 
University of Virginia, University of Washington, and Yale University. 

%    \end{acknowledgements}

%   \bibliography{ref} % your references Yourfile.bib

\bibliographystyle{aa}
\bibliography{bgwmorf}

\begin{appendix}

\section{Minkowski functionals and shapefinders} 
\label{sect:MF}

For a given surface the four Minkowski functionals (from the first to the
fourth) are proportional to the enclosed volume $V$, the area of the surface
$S$, the integrated mean curvature $C$, and the integrated Gaussian curvature
$\chi$. 
Consider an
excursion set $F_{\phi_0}$ of a field $\phi(\mathbf{x})$ (the set
of all points where the density is higher than a given limit,
$\phi(\mathbf{x}\ge\phi_0$)). Then, the first
Minkowski functional is the volume of this region (the excursion set):
\begin{equation}
\label{mf0}
V_0(\phi_0)=\int_{F_{\phi_0}}\mathrm{d}^3x\;.
\end{equation}
The second Minkowski functional is proportional to the surface area
of the boundary $\delta F_\phi$ of the excursion set:
\begin{equation}
\label{mf1}
V_1(\phi_0)=\frac16\int_{\delta F_{\phi_0}}\mathrm{d}S(\mathbf{x})\;.
\end{equation}
The third Minkowski functional is proportional to the
integrated mean curvature $C$ of the boundary:
\begin{equation}
\label{mf2}
V_2(\phi_0)=\frac1{6\pi}\int_{\delta F_{\phi_0}}
    \left(\frac1{R_1(\mathbf{x})}+\frac1{R_2(\mathbf{x})}\right)\mathrm{d}S(\mathbf{x})\;,
\end{equation}
where $R_1(\mathbf{x})$ and $R_2(\mathbf{x})$ 
are the principal radii of curvature of the boundary.

\citet{sah98} and \citet{sss04} used the first three 
Minkowski functionals to define the shapefinders
called as the thickness, the width, and the length as follows. 
The thickness $H_1 = 3V/S$,
the width $H_2 = S/C$, and the length $H_3 = C/4\pi$.  The shapefinders have
dimensions of length and are normalised to give $H_i = R$ for a sphere
of radius $R$.  For smooth (ellipsoidal) surfaces, the shapefinders $H_i$
follow the inequalities $H_1\leq H_2\leq H_3$.  Oblate ellipsoids (pancakes)
are characterised by $H_1 << H_2 \approx H_3$, while prolate ellipsoids
(filaments) are described by $H_1 \approx H_2 << H_3$.

\citet{sah98} also defined  two dimensionless
shapefinders $K_1$ (planarity) and $K_2$ (filamentarity): 
$K_1 = (H_2 - H_1)/(H_2 + H_1)$ and $K_2 = (H_3 -
H_2)/(H_3 + H_2)$. We use these shapefinders in our study
to analyse the shape of the superclusters.

The fourth Minkowski functional is proportional to the integrated
Gaussian curvature (the Euler characteristic) 
of the boundary:
\begin{equation}
\label{mf3}
V_3(\phi_0)=\frac1{4\pi}\int_{\delta F_{\phi_0}}
    \frac1{R_1(\mathbf{x})R_2(\mathbf{x})}\mathrm{d}S(\mathbf{x})\;.
\end{equation}
This functional  describes the topology of the surface; 
it is a sum of the number of isolated clumps 
and the number of void bubbles minus the
number of tunnels (voids open from both sides) in the region
\citep[see, e.g.][]{2002sgd..book.....M,saar06}:

\begin{equation}
\label{v3}
V_3=N_{\mathrm{clumps}} + N_{\mathrm{cavities}} - N_{\mathrm{tunnels}}.
\end{equation}

  High values of the fourth Minkowski
functional $V_3$ suggest a complicated (clumpy) morphology of a
supercluster.

\end{appendix}
\end{document}